\documentclass{PoS}
\usepackage{amsmath}
\title{Time-dependent CP violation in $B$ decays at Belle}

\ShortTitle{Time-dependent CP violation in $B$ decays at Belle}

\author{\speaker{Luka Santelj}\thanks{On behalf of the Belle Collaboration.}\\
        Jozef Stefan Institute, Ljubljana, Slovenia\\
        E-mail: \email{luka.santelj@ijs.si}}

\abstract{Using the full data sample collected with the Belle detector at the KEKB asymmetric-energy $e^+e^-$ collider, we present three recent measurements of time-dependent CP violation in $B$ decays, and a measurement of branching fraction of the $B^0\to\rho^0\rho^0$ decay. We studied $B\to\omega K$ decays and measured the values of CP violation parameters in $B^0\to\omega K^0_S$ to be $A_{\omega K^0_S} =−0.36\pm 0.19 (stat)\pm 0.05 (syst)$ and $S_{\omega K^0_S}= +0.91\pm 0.32 (stat)\pm 0.05 (syst)$, which gives the first evidence of CP violation in this decay. In addition, we measured the direct CP violation in $B^+\to\omega K^+$ to be $A_{CP} (B^+ \to \omega K^+)=−0.03\pm 0.04 (stat)\pm 0.01 (syst)$, and two branching fractions $B(B^0 \to \omega K^0)=(4.5\pm 0.4 (stat)\pm 0.3 (syst)) \times 10^{-6}$ and $B(B^+ \to \omega K^+)=(6.8\pm 0.4 (stat)\pm 0.4 (syst)) \times 10^{-6}$ (preliminary). From the measurement of CP violation parameters in the $B^0\to\eta'K^0$ decay we obtain $S_{\eta'K^0} = 0.68 \pm 0.07(stat)\pm 0.03(syst)$ and $A_{\eta'K^0} = +0.03 \pm 0.05(stat)\pm 0.04(syst)$ (preliminary), which are the world's most precise values to date. Measuring CP violating parameters in the $B^0\to\pi^+\pi^-$ decay gives $A_{\pi^+\pi^-} = +0.33\pm 0.06(stat)\pm 0.03(syst)$ and $S_{\pi^+\pi^-} = -0.64\pm 0.08(stat)\pm 0.03(syst)$. This result is used in an isospin analysis to constrain the $\phi_2$ angle of the unitarity triangle, with which we rule out the region $23.8^\circ < \phi_2 < 66.8^\circ$ at the $1\sigma$ confidence level. The measured branching fraction of the $B^0\to\rho^0\rho^0$ decay is $B(B^0\to\rho^0\rho^0) = (1.02\pm 0.30(stat)\pm 0.15(syst))\times 10^{-6}$, with the fraction of longitudially polarized $\rho^0$ mesons being  $f_L = 0.21^{+0.18}_{-0.22}\pm 0.13$. From the same measurement we obtain also the first evidence of the $B^0\to f_0\rho^0$ decay, by measuring $B(B^0\to f_0 \rho^0)\times B(f_0\to \pi^+\pi^-) = (0.86\pm 0.27(stat) \pm 0.14(syst))\times 10^{-6}$. Using this result in an isospin analysis we obtain $\phi_2 = (91.0\pm 7.2)^\circ$.}

\FullConference{The European Physical Society Conference on High Energy Physics -EPS-HEP2013\\
                18-24 July 2013\\
		Stockholm, Sweden}

\begin{document}

\section{Introduction}

Measuring the parameters of the unitarity triangle (UT) provides a major test of the Standard Model (SM), in particular of the Cabibbo-Kobayashi-Maskawa (CKM) description of flavor changing currents and CP violation. Angles of the UT related to $B_{u,d}$ decays can be determined by measuring CP asymmetries in various $B$ meson decays, and this was the main motivation for construction of two so-called B factory experiments, Belle and BaBar. In the previous decade both experiments have confirmed the complex phase of the CKM matrix as the main source of CP violation.   

In these proceedings we present two recent measurements related to the $\phi_1$ angle of the UT, mainly motivated by their sensitivity to possible New Physics contributions, and two measurements of the $\phi_2$ angle. All measurements are based on the data sample containing 772 millions $B\bar{B}$ pairs collected by the Belle experiment \cite{belle} , during its full data taking period (1999-2010). 

Angles $\phi_1$ and $\phi_2$ can be determined by measuring time-dependent asymmetry between decays of $B^0$ and $\bar{B}^0$ mesons into a common CP eigenstate $f_{CP}$ \cite{bigi}. At the Belle experiment pairs of $B$ mesons are produced in asymmetric energy collisions of electrons and positrons, through the $\Upsilon(4S)\to B_{tag}B_{CP} \to f_{tag}f_{CP}$ process. Since a $B$ meson pair is in a quantum coherent state, a decay of $B_{tag}$ into a flavor specific final state $f_{tag}$ at $t_{tag}$, determines the flavor of $B_{CP}$ at $t_{tag}$. In this case the CP asymmetry is given by\footnote{Here $B^0(t)$ ($\bar{B}^0(t)$) denote states that were at $t=0$ pure $B^0$ ($\bar{B}^0$) states, but later get mixed due to $B^0-\bar{B}^0$ mixing.}  
\begin{equation}
a_{CP}(\Delta t) = \frac{\Gamma(B^0(\Delta t)\to f_{CP}) - \Gamma(\bar{B}^0(\Delta t)\to f_{CP}) }{\Gamma(B^0(\Delta t)\to f_{CP}) + \Gamma(\bar{B}^0(\Delta t)\to f_{CP})} = A_{f}\cos\Delta M \Delta t + S_{f}\sin \Delta M \Delta t,
\end{equation}
where $\Delta t$ is the time difference between decays of $B_{tag}$ and $B_{CP}$, $\Delta M$ is the mass difference between the two $B^0$ mass eigenstates ($B_L$ and $B_H$), and $A_f$ and $S_f$ are the so-called CP violation parameters, which can be within the SM related with the UT angles.  

\section{Measurement of branching fractions and CP violation parameters in $B\to\omega K$ decays}
The $B^0\to \omega K^0_S$ decays are sensitive to the $\phi_1= \arg(-V_{cd}V^*_{cb})/(V_{td}V^*_{tb})$ interior angle of the UT. The decay proceeds dominantly by the $b\to s\bar{q}q$ penguin diagram, and within the SM we expect $A_{\omega K^0_S} = 0$ and $S_{\omega K^0_S} = \sin 2\phi_1$, neglecting other contributing CKM-suppressed amplitudes with a different weak phase. However, the contribution of these CKM-suppressed amplitudes may not be negligible, resulting in a non-zero $A_{\omega K^0_S}$ and in a deviation of $S_{\omega K^0_S}$ from $\sin 2\phi_1$. Several theoretical methods were used to estimate the effect of these amplitudes, indicating the expected value of $S_{\omega K^0_S}$ slightly higher than $\sin 2\phi_1$ \cite{b2s}. However, current experimental measurements indicate the opposite \cite{belle_omega,babar_omega,hfag}, which might be a consequence of a contribution of new heavy particles in the loop of the penguin diagram \cite{npcont}. 

In this measurement, we have also measured the direct CP violating parameter $A_{CP}$ in the $B^+\to \omega K^+$ decay, defined as
\begin{equation}
A_{CP} =\frac{\Gamma(B^-\to\omega K^-) - \Gamma(B^+\to\omega K^+) }{\Gamma(B^-\to\omega K^-) + \Gamma(B^+\to\omega K^+)},
\end{equation}
where again a deviation from the expected asymmetry could be an indication of New Physics. Furthermore, the measurement of the branching fractions provides an important test of the QCD factorization (QCDF) and perturbative QCD (pQCD) approaches.

\begin{figure}[ht!]
\centering
  \includegraphics[width=14cm]{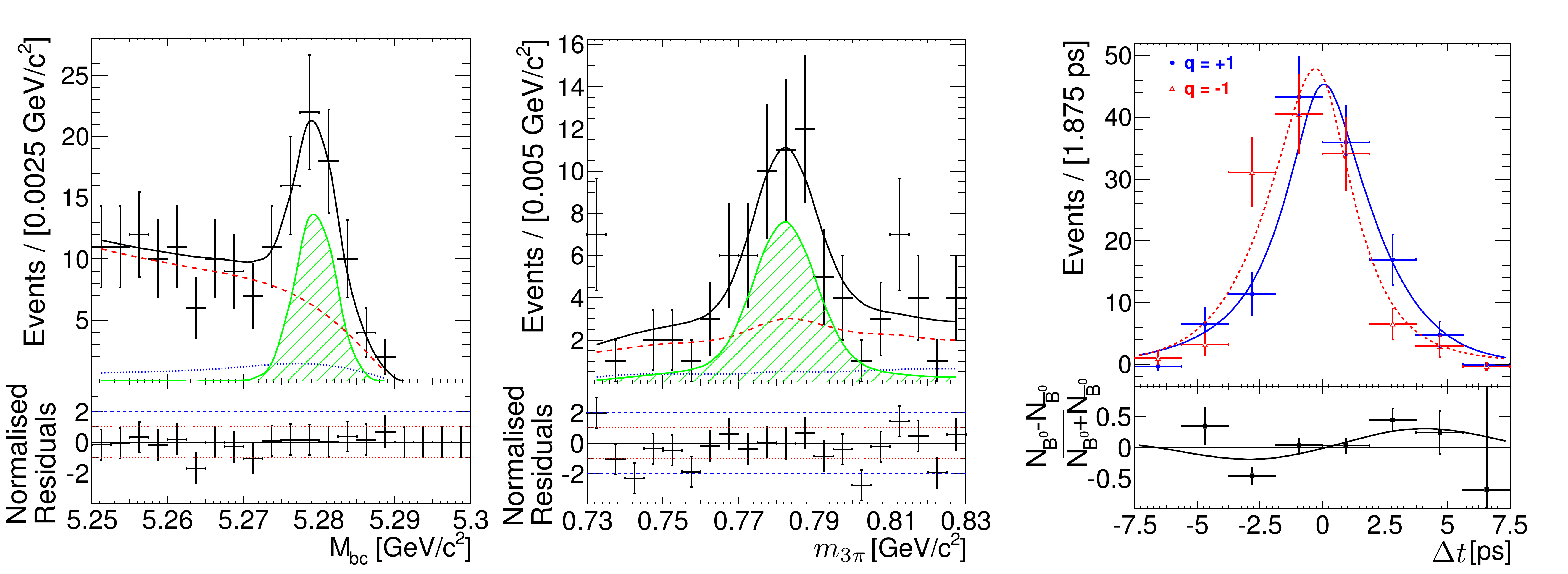}
  \caption{{\bf Left two:} Distribution of reconstructed events in $M_{bc}$ and $m_{3\pi}$ (black points) along with the fitted PDF (the full line). The dashed line shows contribution of the $q\bar{q}$ background (from $e^+e^-\to q\bar{q}~ (q=u,d,s,c)$ events), the dotted line of the $B\bar{B}$ background. {\bf Right:} Distributions of reconstructed events in $\Delta t$ (events with $q=+1$ (blue points) and $q=-1$ (red points)) along with the corresponding $q=+1$ and $q=-1$ parts of the fitted PDF; bottom plot shows the asymmetry in data distributions and in the fitted PDF.}
\label{fig:omega}
\end{figure}

To obtain the two branching fractions and CP violation parameters we perform a seven-dimensional unbinned extended maximum likelihood fit to $M_{bc},\Delta E$ (two kinematic variables of the reconstructed B meson), $R_{s/b}$ (event topology variable), $m_{3\pi}$ (invariant mass of the reconstructed $\omega$), $H_{3\pi}$ (helicity angle), $\Delta t$ and $q$ (where $q=+1$ ($q=-1$) for $B_{tag}=B^0$ ($\bar{B}^0$)). The fit is performed simultaneously to $B^0\to\omega K^0_S$ and $B^+\to \omega K^+$ data samples, sharing common calibration factors. Following this, the model shape is fixed and the $A_{CP} (B^+ \to\omega K^+)$ parameter is obtained from two further fits to extract the number of $B^+$ and $B^-$ events. The preliminary results are \cite{omegaks}:
\begin{align}
B(B^0 \to \omega K^0)&=(4.5\pm 0.4 (stat)\pm 0.3 (syst)) \times 10^{-6}, \nonumber \\
B(B^+ \to \omega K^+)&=(6.8\pm 0.4 (stat)\pm 0.4 (syst)) \times 10^{-6},\nonumber \\
A_{\omega K^0_S}&=−0.36\pm 0.19 (stat)\pm 0.05 (syst),\nonumber \\
S_{\omega K^0_S}&= +0.91\pm 0.32 (stat)\pm 0.05 (syst), \nonumber\\
A_{CP} (B^+ \to \omega K^+)&=−0.03\pm 0.04 (stat)\pm 0.01 (syst), 
\label{omegak}
\end{align}
where the first uncertainty is statistical and the second is systematic. The latter is dominated by uncertainties of the $\Delta t$ resolution function parameters for $A_{\omega K^0_S}$ and $S_{K^0_S}$, and by parameters of the background PDF shape for the branching fractions. The comparison of data distributions and the fitted PDF is shown in figure \ref{fig:omega}. The results given in (\ref{omegak}) are the world's most precise measurements of the branching fractions and CP violation parameters in $B\to\omega K$ decays. The observed values of $A_{\omega K^0_S}$ and $S_{\omega K^0_S}$ differ from zero with a significance of 3.1 standard deviations, which gives the first evidence of CP violation in the $B^0\to\omega K^0_S$ decay.

\section{Measurement of CP violation parameters in $B^0\to\eta'K^0$ decay}

The $B^0\to\eta'K^0$ decay is another decay that proceeds dominantly through the $b\to s\bar{q}q$ transition, for which within the SM we expect $A_{\eta'K^0}=0$ and $S_{\eta'K^0}=-\xi_f\sin 2\phi$ (where $\xi_f=-1~(+1)$ for $B\to\eta'K^0_S$ ($B\to\eta'K^0_L$)). Theoretically this is the cleanest mode to measure CP violation parameters in a $b\to s\bar{q}q$ process, as the contributions from the CKM suppressed diagrams are expected to be $\lesssim 0.02$ for both $S_{\eta K^0}$ and $A_{\eta'K^0}$ \cite{b2s}. 

\begin{figure}[ht!]
  \centering
  \includegraphics[width=14cm]{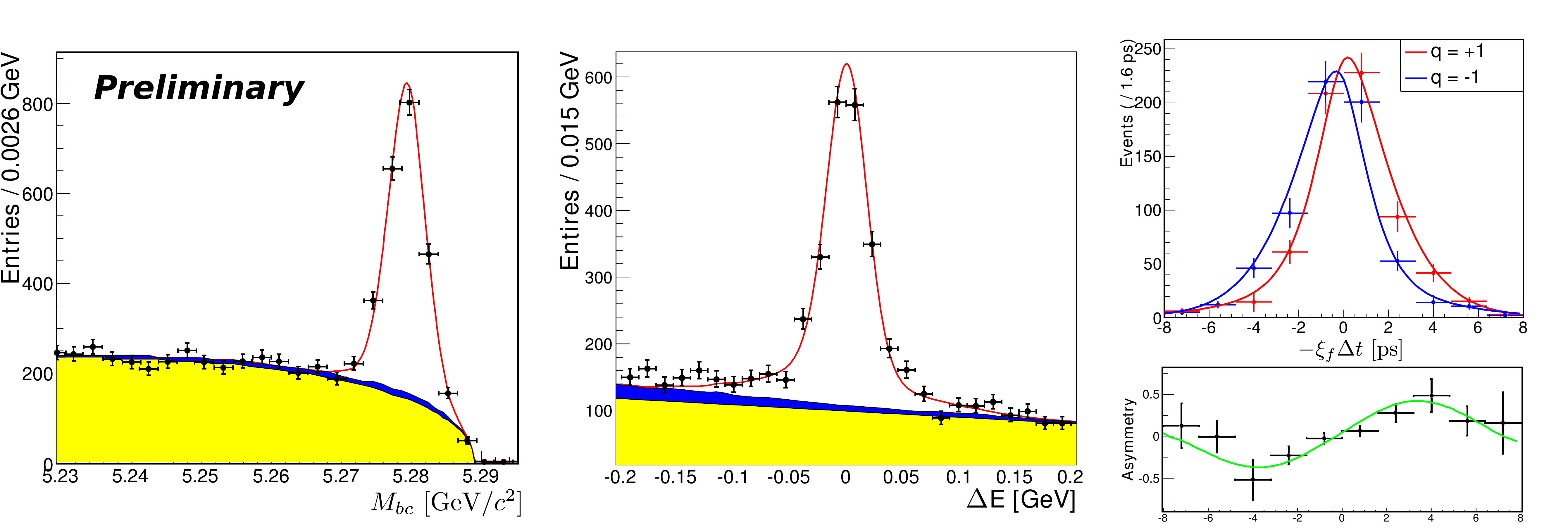}
  \caption{{\bf Left two:} Distribution of reconstructed $B^0\to\eta'K^0_S$ events in $M_{bc}$ and $\Delta E$ (black points) along with the fitted PDF (the red). The yellow area shows contribution of the $q\bar{q}$ background, and the blue area of the $B\bar{B}$ background. {\bf Right:} Distribution of the reconstructed events in $\Delta t$ (events with $q=+1$ (red points) and $q=-1$ (blue points))  along with the corresponding $q=+1$ and $q=-1$ parts of the fitted PDF; bottom plot shows the asymmetry in data distributions and in the fitted PDF.}
\label{fig:etap}
\end{figure}

In the first part of the analysis, event reconstruction and signal fraction estimation, we study separately $B^0\to \eta'K^0_S$ and  $B^0\to \eta'K^0_L$ events. To obtain the fraction of signal events we study the distribution of events in $M_{bc},\Delta E$ and $R_{s/b}$ for $K^0_S$ events, and the distribution in $p_B^{cms}$ ($B$ candidate momentum in the center-of-mass system), $r$ (quality of the B candidate flavor information) and $R_{s/b}$ for $K^0_L$ events. Altogether we reconstruct $2503\pm 63$ $B^0\to\eta'K^0_S$ signal events, and $1041\pm 41$ $B^0\to\eta'K^0_L$ signal events, where the uncertainties are statistical only. Following this, we perform an unbinned maximum likelihood fit to extract the values of CP violation parameters from the measured $\Delta t,q$ distribution of events.  Our preliminary results are 
\begin{align}
S_{\eta'K^0} &= +0.68 \pm 0.07(stat)\pm 0.03(syst),\nonumber \\
A_{\eta'K^0} &= +0.03 \pm 0.05(stat)\pm 0.04(syst),
\label{drek}
\end{align}
where the first uncertainty is statistical and the second is systematic. The main contribution to the latter comes from the uncertainties in the $\Delta t$ resolution function parameters for $S_{\eta'K^0}$, and from the tag-side interference effect for $A_{\eta'K^0}$. The comparison of data distribution and the fitted PDF is shown in figure \ref{fig:etap}. The measured values of $S_{\eta'K^0}$ and $A_{\eta'K^0}$ are the world's most precise values of CP violation parameters in this particular decay, as well as among all $b\to s\bar{q}q$ transition dominated decays. They are consistent with previous measurements \cite{belle_etap,babar_etap} and with the SM prediction.  

\newpage
\section{Measurements of $\phi_2$ angle in $B^0 \to\pi^+\pi^-$ and $B^0\to \rho^0\rho^0$ decays}

Decays  $B \to\pi^+\pi^-$ and $B\to \rho^0\rho^0$ are sensitive to $\phi_1= \arg(-V_{td}V^*_{tb})/(V_{ud}V^*_{ub})$. At the tree level we expect $A_{f}=0$ and $S_{f}=\sin 2\phi_2$. However, penguin contributions can give rise to direct CP violation, $A_{f}\neq 0$, and also pollute the measurement of $\phi_2$. Despite this, it is still possible to obtain the value of $\phi_2$ with a $SU(2)$ isospin analysis, by considering the set of three $B\to hh$ decays ($h=\pi$ or $h=\rho$), related via isospin symmetry \cite{isospin}. Belle recently updated measurements of CP violation parameters in the $B^0\to\pi^+\pi^-$ decay and of branching fraction of the $B^0\to\rho^0\rho^0$ decay to the full data sample, and the values obtained were used to provide new constraints on $\phi_2$. 

\subsection{The $B^0\to\pi^+\pi^-$ decay}

\begin{figure}[ht!]
 \includegraphics[width=6.5cm]{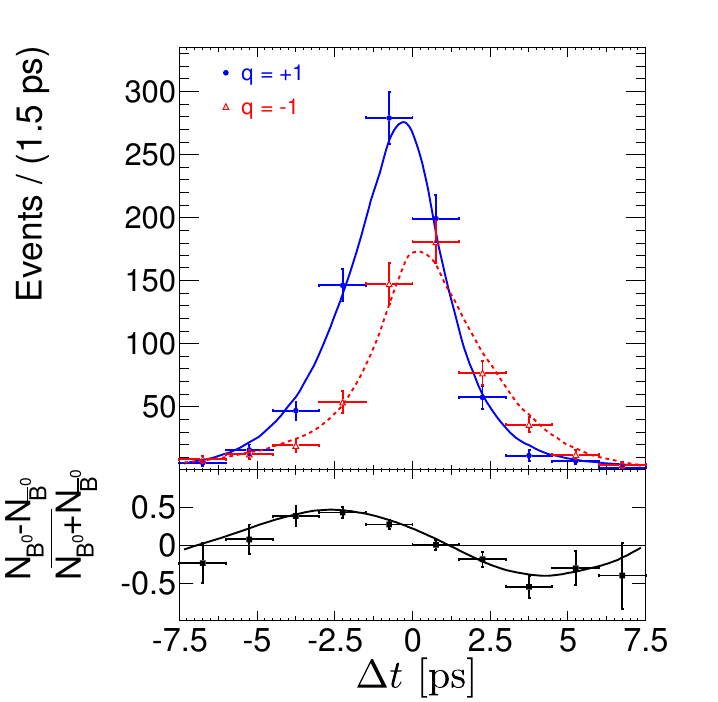}
 \includegraphics[width=7cm]{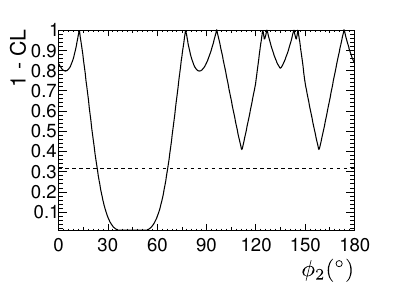}
\caption{{\bf Left:} Distribution of reconstructed events in $\Delta t$ (events with $q=+1$ (red points) and $q=-1$ (blue points))  along with the corresponding $q=+1$ and $q=-1$ parts of the fitted PDF. {\bf Right:} Difference 1-CL, plotted for a range of $\phi_2$. The dashed line indicate the $1\sigma$ exclusion level.}
\label{fig:pipi}. 
\end{figure}

To obtain the values of $S_{\pi^+\pi^-}$ and $A_{\pi^+\pi^-}$ a seven-dimensional fit to $L^+_{K\pi},L^-_{K\pi}$ ($\pi,K$ separation likelihood function), $M_{bc}, \Delta E,R_{s/b},\Delta t$ and $q$ is performed. The main background contribution comes from $e^+e^- \to q\bar{q}~(q=u,d,s,c)$ events. In total we reconstruct $2964\pm 88$ signal events, and the following values of CP violation parameters are obtained \cite{pipi}
\begin{align}
A_{\pi^+\pi^-} = +0.33\pm 0.06(stat)\pm 0.03(syst), \nonumber \\
S_{\pi^+\pi^-} = -0.64\pm 0.08(stat)\pm 0.03(syst),
\label{pipi}
\end{align}
where the first uncertainty is statistical and the second is systematic. The comparison of data distribution in $\Delta t$ with the fitted PDF is shown in figure \ref{fig:pipi}. The values given in (\ref{pipi}) are the world's most precise values of CP violation parameters in this decay. Using these values, and input from other Belle measurements (branching fractions of $B^0\to\pi^+\pi^-$, $B^+\to \pi^+\pi^0$ \cite{pipi_cite} and $B^0\to\pi^0 \pi^0$ \cite{pipi_cite1} decays), an isospin analysis is performed to constrain $\phi_2$. Obtained difference 1-CL (confidence level) is plotted in figure \ref{fig:pipi}, for a range of $\phi_2$. The region $23.8^\circ < \phi_2 < 66.8^\circ$ is ruled out at the $1\sigma$ level, including systematic uncertainties.

\subsection{The $B^0\to\rho^0\rho^0$ decay}

Measuring the branching fraction of the $B^0\to\rho^0\rho^0$ decay is quite challenging due to its low value and the presence of other, largely unknown, four-pion final states. In addition, due to two vector particles in the  final state, a helicity analysis is needed in order to separate the longitudinal and transverse polarization amplitudes, with even and odd CP eigenvalues respectively.     

We perform an unbinned maximum likelihood fit to a six-dimensional distribution of reconstructed candidates in $m^1_{\pi^+\pi^-}, m^2_{\pi^+\pi^-}$ (invariant masses of reconstructed $\rho^0s$), $\cos\theta^1_{hel},\cos\theta^2_{hel}$ (helicity angles), and $R_{s/b}$. The results are \cite{rho}
\begin{align} 
  B(B^0\to\rho^0\rho^0) &= (1.02\pm 0.30(stat)\pm 0.15(syst))\times 10^{-6}, \nonumber \\
  f_L &= 0.21^{+0.18}_{-0.22}(stat)\pm 0.13(syst), 
\end{align}
where B is a branching fraction and $f_L$ is a fraction of longitudinally polarized $\rho^0$ mesons. The branching fraction is measured with a significance of $3.4$ standard deviations. In addition, we measured 
\begin{equation}
B(B^0\to f_0 \rho^0)\times B(f_0\to \pi^+\pi^-) = (0.86\pm 0.27(stat) \pm 0.15(syst))\times 10^{-6},
\end{equation}
with a significance of $3.0$ standard deviations, which is the first evidence of the $B^0\to f_0 \rho^0$ decay.

Using the values of $B(B^0\to\rho^0\rho^0)$ and $f_L$, along with the world average values \cite{hfag} of $B(B^0\to\rho^+\rho^-),~f^{+-}_L,~A_{\rho^+\rho^-},~S_{\rho^+\rho^-}, ~B(B^+\to\rho^+\rho^0)$ and CP violation parameters $A_{\rho^0\rho^0},S_{\rho^0\rho^0}$ from BaBar measurement \cite{babar_rho0}, an isospin analysis is performed to constrain $\phi_2$. We obtain $\phi_2 = (91.0 \pm 7.2)^\circ$, at a $1\sigma$ confidence level.

\end{document}